\documentclass[
prb,
twocolumn,
dvips,
aps]{revtex4}

\usepackage{amsmath,amssymb,amsfonts,graphicx}

\begin{document}

\title{Heat transfer between nanoparticles: Thermal conductance for near-field interactions}
\author{A. P\'{e}rez-Madrid}
\email[E-mail:~]{agustiperezmadrid@ub.edu}
\affiliation{Departament de F\'{\i}sica Fonamental, Facultat de F\'{\i}sica, Universitat
de Barcelona, Av. Diagonal 647, 08028 Barcelona, Spain}
\author{J. M. Rub\'{\i}}
\email[E-mail:~]{mrubi@ub.edu}
\affiliation{Departament de F\'{\i}sica Fonamental, Facultat de F\'{\i}sica, Universitat
de Barcelona, Av. Diagonal 647, 08028 Barcelona, Spain}
\author{L. C. Lapas}
\email[E-mail:~]{luciano@fis.unb.br}
\affiliation{Instituto de F\'{\i}sica and Centro Internacional de F\'{\i}sica da Mat\'{e}%
ria Condensada, Universidade de Bras\'{\i}lia, Caixa Postal 04513, 70919-970
Bras\'{\i}lia, Distrito Federal, Brazil}
\pacs{PACS number}

\begin{abstract}
We analyze the heat transfer between two nanoparticles separated by a
distance lying in the near-field domain in which energy interchange is due
to Coulomb interactions. The thermal conductance is computed by assuming
that the particles have charge distributions characterized by fluctuating
multipole moments in equilibrium with heat baths at two different
temperatures. This quantity follows from the fluctuation-dissipation theorem
(FDT) for the fluctuations of the multipolar moments. We compare the
behavior of the conductance as a function of the distance between the
particles with the result obtained by means of molecular dynamics
simulations. The formalism proposed enables us to provide a comprehensive
explanation of the marked growth of the conductance when decreasing the
distance between the nanoparticles.
\end{abstract}

\startpage{1}
\endpage{102}
\maketitle

\section{Introduction}

The study of energy transfer mechanisms at the nanoscale~\cite%
{volz,greffet3} has aroused increasing interest due to the emergence of the
interdisciplinary field of nanoscience where such wide-ranging fields as for
example solid state physics~\cite{greffet5}, nanothermodynamics~\cite%
{rubi,rubi2,hill} or electrical engineering~\cite{volokitin} coexist. One of
the basic problems in this field is to determine the energy exchange between
two nanoparticles (NPs) at different temperatures. The way in which this
energy is transfered depends crucially on the distance between the
particles. For sufficiently large distances, heat exchange proceeds via
thermal radiation, through emission or absorption of photons whereas at
smaller distances recent molecular dynamics simulations have shown that
Coulomb interaction (near-field radiation) is the dominant mechanism~\cite%
{greffet}.

For near-field interactions, the thermal conductance was calculated under
the assumption that both NPs behave as effective dipoles at different
temperatures~\cite{greffet}. Hence, since these dipoles undergo thermal
fluctuations, the fluctuation-dissipation theorem (FDT)~\cite%
{kubo,callen,landau,Groot} provides the energy which dissipates into heat in
each NP. It was found that the heat and therefore the conductance varies
according to $d^{-6}$ a very different behavior from the one observed in the
case of thermal radiation: $d^{-2}$. Molecular dynamics simulations agree
with the dipole-dipole model when the two NPs are separated by a distance on
the order of a few nanometers. However, near contact the conductance
deviates dramatically from the prediction of the dipole model, as the
simulations show. This behavior is a consequence of\textbf{\ } the fact that
when particles become very close the position of the atoms are highly
correlated, consequently the charge distributions\textbf{\ }become
nonsymmetric and cannot be described merely as two interacting dipoles. To
account for this distortion of the distribution of charges a more general
formalism which focuses more convoluted interactions involving higher order
multipoles aside from the dipoles is required.

Our purpose in this paper is to provide this general formalism enabling us
to analyze the behavior of the conductance beyond the dipolar approximation.
We will use the linear response theory to derive an expression of the FDT
for the fluctuations of the higher order multipoles. In particular, we will
focus on the quadrupolar contributions to the conductance which are able to
reproduce the behavior observed in the simulations for some sizes of the NPs.

The paper is organized as follows. In Section 2, we present the multipolar
expansion of the Coulomb forces~\cite{stone} between both NPs and derive a
general expression of the FDT valid for multipoles of any order which leads
to the heat transfer between the NPs. In Section 3, we analyze the
particular case of quadrupolar contributions and derive the expresion of the
conductance. We compare our result with the molecular dynamics simulations~%
\cite{greffet}. Finally, in Section 4, we emphasize our main conclusions.

\section{Heat transfer between two nanoparticles}

In this section, we will study the near-filed radiative heat transfer flux
between two NPs which interact through Coulomb forces.

\subsection{Multipolar expansion}

\begin{figure}[tbp]
\includegraphics[scale=1.0]{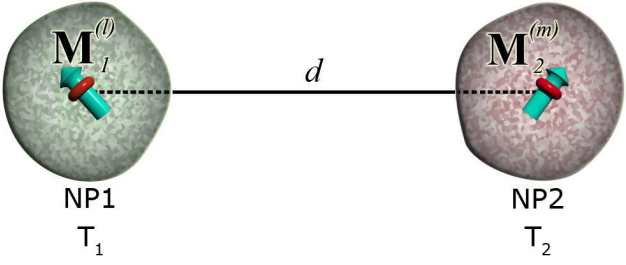}
\caption{Schematic illustration of the interaction between two NPs (NP1 and
NP2) with temperatures $T_{1}$ and $T_{2}$, respectively. Each NP is
assimilated to a multipole moment (moments $\mathbf{M}_{1}^{(l)}$ and $%
\mathbf{M}_{2}^{(m)}$), and are separated by a distance $d$ between their
centers.}
\label{figure_NPs}
\end{figure}

To analyze the Coulomb interaction between two NPs (see Fig.~\ref{figure_NPs}%
) it is necessary to know the charge distribution inside each of them. This
can be performed by specifying their multipole moments so that the multipole
moment of order $n$ of the NP$i$, $\mathbf{\hat{M}}_{(i)}^{(n)}$, can be
defined as~\cite{hess} 
\begin{equation}
\hat{M}_{(i);\alpha }^{(n)}(\mathbf{r})=\frac{1}{n!}\sum_{\mathbf{r}}e_{%
\mathbf{r}}r^{2n+1}X_{\alpha }^{(n)}(\mathbf{r}),  \label{multipole}
\end{equation}%
where $e_{\mathbf{r}}$ is the charge at the position $\mathbf{r}$ inside the
NP and the $X_{\alpha }^{(n)}(\mathbf{r})$ are symmetric irreducible tensors
(see the Appendix I for more details) where $\alpha =\left( \alpha
_{1},\ldots ,\alpha _{n}\right) $ and $\alpha _{j}=1,2,3$ for $j=1,...,n$.
Thus, the case $n=0$ corresponds to the monopole $\hat{M}_{(i);\alpha
}^{(0)}(\mathbf{r})=\sum_{\mathbf{r}}e_{\mathbf{r}}$, $n=1$ is related to
the dipole moment $\hat{M}_{(i);\alpha }^{(1)}(\mathbf{r})=\sum_{\mathbf{r}%
}e_{\mathbf{r}}r_{\alpha _{1}}$, and $n=2$ for the quadrupole moment $\hat{M}%
_{(i);\alpha }^{(2)}(\mathbf{r})=1/2\sum_{\mathbf{r}}e_{\mathbf{r}}\left(
3r_{\alpha _{1}\alpha _{2}}-\delta _{\alpha _{1}\alpha _{2}}\right) $.
Hence, in terms of the spherical surface tensors $Y_{\alpha }^{(n)}(\mathbf{%
\hat{r}})$, given through Eq. (\ref{surf_tens}), and the unit vector $%
\mathbf{\hat{r}}$ related to $\mathbf{r}$, Eq. (\ref{multipole}) adopts the
equivalent form 
\begin{equation}
\hat{M}_{(i);\alpha }^{(n)}(\mathbf{r})=\frac{1}{n!}\sum_{\mathbf{r}}e_{%
\mathbf{r}}r^{n}Y_{\alpha }^{(n)}(\mathbf{\hat{r}})\text{.}
\label{multipole_2}
\end{equation}

The above mentioned interaction between these NPs modifies their respective
Hamiltonians. The interaction between NP$i$ with NP$j$ introduces a
time-dependent perturbation $\hat{H}_{(ij)}$ in its Hamiltonian which can be
written as a multipolar expansion~\cite{stone}: 
\begin{equation}
\hat{H}_{(ij)}=\sum_{m=0}^{\infty }c_{m}\mathbf{\hat{M}}_{(i)}^{(m)}\odot 
\mathbf{\hat{V}}_{(i,j)}^{(m)}(t)\text{,}  \label{hamiltonian_inter}
\end{equation}%
with $c_{m}=1/(2m-1)!!$ and $\odot$ stands for the full contraction of
indexes, $\mathbf{\hat{M}}_{(i)}^{(m)}\odot \mathbf{\hat{V}}%
_{(i,j)}^{(m)}\equiv \hat{M}_{(i);\alpha }^{(m)}\hat{V}_{(i,j);\alpha
}^{(m)} $. In addition 
\begin{equation}
\hat{V}_{(i,j);\alpha }^{(m)}=\nabla _{\alpha _{1}}\ldots \nabla _{\alpha
_{m}}\mathbf{\hat{V}}_{(i,j)}(d)\text{,}  \label{potential_grad}
\end{equation}%
with $\mathbf{\hat{V}}_{(i,j)}(d)$ being the interaction potential between
both NPs and $d$ the separation between their centers. In terms of the first
contributions, the perturbation can be expressed as 
\begin{eqnarray}
\hat{H}_{(ij)} &=&\mathbf{\hat{M}}_{(i)}^{(0)}\mathbf{\hat{V}}_{(i,j)}+%
\mathbf{\hat{M}}_{(i)}^{(1)}\cdot \mathbf{\hat{V}}_{(i,j)}^{(1)}  \notag \\
&&+\frac{1}{3}\mathbf{\hat{M}}_{(i)}^{(2)}\odot \mathbf{\hat{V}}%
_{(i,j)}^{(2)}+\ldots \text{,}  \label{hamil_2}
\end{eqnarray}%
where $-\mathbf{\hat{V}}_{(i,j)}^{(1)}$ is the electric field induced in the
NP$i$, $-\mathbf{\hat{V}}_{(i,j)}^{(2)}$ is the gradient of this induced
field, and $\mathbf{\hat{M}}_{(i)}^{(2)}$ is the conjugated quadrupolar
moment.

Likewise, the electrostatic potential admits a multipolar expansion as well 
\begin{equation}
\mathbf{\hat{V}}_{(i,j)}(d)=\sum_{n=0}^{\infty}c_{n}\mathbf{G}%
_{(i,j)}^{(n)}\odot\mathbf{\hat{M}}_{j}^{(n)}  \label{multipolar_potential}
\end{equation}
which expresses the fact that the potential acting on the NP$i$ depends on
the charge distribution in the NP$j$. Here, $\mathbf{\hat{M}}_{(j)}^{(n)}$
are the multipolar moments of the NP$j$ and 
\begin{eqnarray}
G_{(i,j);\alpha}^{(n)}&=&\frac{(-1)^{n}}{4\pi\varepsilon_{0}}%
\nabla_{\alpha_{1}}\ldots\nabla_{\alpha_{n}}\frac{1}{d}=\frac{1}{%
4\pi\varepsilon_{0}}X_{\alpha}^{(n)}(\mathbf{d})  \notag \\
&=&\frac{1}{4\pi\varepsilon_{0}d^{n+1}}Y_{\alpha}^{(n)}(\mathbf{\hat{d}})
\label{green}
\end{eqnarray}
is the Green propagator, with $\mathbf{d}$ the vector connection the centers of
the particles and $\mathbf{\hat{d}}$ the corresponding unit vector. Thus, from
Eq. (\ref{multipolar_potential}) 
\begin{equation}
\mathbf{\hat{V}}_{(i,j)}^{(m)}=\sum_{n=0}^{\infty}c_{n}\mathbf{G}%
_{(i,j)}^{(m,n)}\odot\mathbf{\hat{M}}_{j}^{(n)}\text{,}
\label{multipol_rela}
\end{equation}
where $\mathbf{G}_{(i,j)}^{(m,n)}$ is defined through 
\begin{equation}
\mathbf{G}_{(i,j);\beta,\alpha}^{(m,n)}=\nabla_{\beta_{1}}\ldots\nabla_{%
\beta_{m}}G_{(i,j);\alpha}^{(n)}\text{.}  \label{green3}
\end{equation}

\subsection{Heat transfer from the fluctuation-dissipation theorem}

In the linear response regime, the multipolar moments can be expressed as 
\begin{equation}
\mathbf{M}_{(i)}^{(n)}(\omega)=\frac{1}{c_{n}}\sum_{m=0}^{\infty}\mathbf{P}%
_{(i)}^{(n,m)}(\omega)\odot\mathbf{V}_{(i,j)}^{(m)}(\omega)\text{.}
\label{lin_res}
\end{equation}
where $\mathbf{P}^{(n,m)}(\omega)$ are the multipolar polarizabilities which
may in general depend on frequency.

The energy transferred between the particles and converted into heat can be
obtained from the linear response theory~\cite{landau,Groot}. One obtains
(see the Appendix II) 
\begin{eqnarray}
Q_{i\rightarrow j} &=&\frac{-i\omega \varepsilon _{0}}{4}\sum_{n,m=0}^{%
\infty }\mathbf{\{}c_{n}\left\langle \mathbf{V}_{(i,j)}^{(n)\ast }\odot 
\mathbf{P}_{(j)}^{(n,m)}\odot \mathbf{V}_{(i,j)}^{(m)}\right\rangle -  \notag
\\
&&c_{m}\left\langle \mathbf{V}_{(i,j)}^{(m)}\odot \mathbf{P}%
_{(j)}^{(m,n)\ast }\odot \mathbf{V}_{(i,j)}^{(n)\ast }\right\rangle \mathbf{%
\}}\text{,}  \label{energy}
\end{eqnarray}%
where the symbol $^{\ast }$ stands for the complex conjugated, and the
brakets express thermal average.

According to Eq. (\ref{multipol_rela}), the term in Eq. (\ref{energy})
containing the thermal average can be transformed as 
\begin{gather}
\sum_{n,m=0}^{\infty }\left\langle \mathbf{V}_{(i,j)}^{(n)\ast }\odot 
\mathbf{P}_{(j)}^{(n,m)}\odot \mathbf{V}_{(i,j)}^{(m)}\right\rangle
=\sum_{l,k=0}^{\infty }c_{l}c_{k}\sum_{n,m=0}^{\infty }\times  \notag \\
\left\langle \mathbf{M}_{(i)}^{(l)\ast }\odot \mathbf{G}^{(l,n)}\odot 
\mathbf{P}_{(j)}^{(n,m)}\odot \mathbf{G}^{(m+k)\mathbf{\ }}\odot \mathbf{M}%
_{i}^{(k)}\right\rangle =  \notag \\
\sum_{l,k=0}^{\infty }c_{l}c_{k}\left\langle \mathbf{M}_{(i)}^{(l)\ast
}\odot \mathbf{S}_{(j)}^{(l,k)}\odot \mathbf{M}_{i}^{(k)}\right\rangle \text{%
,}  \label{correlation_1}
\end{gather}%
where we have defined 
\begin{equation}
\mathbf{S}_{(j)}^{(l,k)}=\sum_{n,m=0}^{\infty }\mathbf{G}^{(l,n)}\odot 
\mathbf{P}_{(j)}^{(n,m)}\odot \mathbf{G}^{(m,k)\mathbf{\ }}\text{.}
\label{kernel}
\end{equation}%
Moreover, from Eqs. (\ref{surf_tens}), (\ref{green}) and (\ref{green3}) one
can prove that the $\mathbf{S}_{(j)}^{(l,k)}$ are symmetric tensors.
Therefore, making use of Eq. (\ref{kernel}), Eq. (\ref{energy}) becomes 
\begin{eqnarray}
Q_{i\rightarrow j} &=&\frac{-i\omega \varepsilon _{0}}{4}\sum_{l,k=0}^{%
\infty }c_{l}c_{k}\times  \notag \\
&&\left\{ \left\langle \mathbf{M}_{(i)}^{(l)\ast }\odot \mathbf{S}%
_{(j)}^{(l,k)}\odot \mathbf{M}_{i}^{(r)}\right\rangle -\text{c.c.}\right\} 
\text{.}  \label{energy_2}
\end{eqnarray}%
The dependence of the energy transferred on the distance $d$ resides in $%
\mathbf{S}_{(j)}^{(l,k)}$, as follows from Eq. (\ref{kernel}) and the
expression of the propagators given through Eqs. (\ref{green}) and (\ref%
{green3}). The multipole-multipole correlation can be obtained by using the
FDT~\cite{landau,Groot,callen} 
\begin{equation}
\left\langle \mathbf{M}_{(i)}^{(l)\ast }\mathbf{M}_{(i)}^{(k)}\right\rangle =%
\frac{-i\varepsilon _{0}}{\pi \omega c_{l}c_{k}}\left( \mathbf{P}%
_{(i)}^{(l,k)}-\mathbf{P}_{(i)}^{(k,l)\ast }\right) \Theta (\omega ,T_{i})%
\text{,}  \label{FD}
\end{equation}%
where $\Theta (\omega ,T_{i})=\hbar \omega \left\{ 1/2+1/\exp (\hbar \omega
/kT_{i}-1)\right\} $ is the mean energy of an oscillator. As an
illustration, for the dipolar case~\cite{greffet}, we obtain 
\begin{equation}
\left\langle \mathbf{M}_{(i)}^{(1)\ast }\mathbf{M}_{(i)}^{(1)}\right\rangle =%
\frac{-i\varepsilon _{0}}{\pi \omega }\left( \mathbf{P}_{(i)}^{(1,1)}-%
\mathbf{P}_{(i)}^{(1,1)\ast }\right) \Theta (\omega ,T_{i})\text{,}
\label{FD_dip}
\end{equation}%
where $\mathbf{P}_{(i)}^{(1,1)}$ is the dipole-dipole polarizability which
we assume to be given through 
\begin{equation}
P_{(i);\alpha ,\beta }^{(1,1)}=\alpha _{(i)}(\omega )\bigtriangleup _{\alpha,
\beta }^{(1)}  \label{P_1_1}
\end{equation}%
and $\bigtriangleup _{\alpha , \beta }^{(1)}=\delta _{\alpha _{1}\beta _{1}}$,
with $\alpha _{(i)}(\omega )=\alpha _{(i)}^{\shortmid }(\omega )+i\alpha
_{(i)}^{\shortparallel }(\omega )$. Hence, Eq. (\ref{FD_dip}) becomes 
\begin{equation}
\left\langle M_{(i);\alpha }^{(1)\ast }M_{(i);\beta }^{(1)}\right\rangle =%
\frac{2\varepsilon _{0}}{\pi \omega }\alpha _{(i)}^{\shortparallel }(\omega
)\Theta (\omega ,T_{i})\delta _{\alpha _{1}\beta _{1}}\text{.}
\label{FD_dip_2}
\end{equation}%
For the quadrupolar case one has 
\begin{equation}
\left\langle \mathbf{M}_{(i)}^{(2)\ast }\mathbf{M}_{(i)}^{(2)}\right\rangle =%
\frac{-i\varepsilon _{0}}{9\pi \omega }\left( \mathbf{P}_{(i)}^{(2,2)}-%
\mathbf{P}_{(i)}^{(2,2)\ast }\right) \Theta (\omega ,T_{i})\text{,}
\label{FD_qua}
\end{equation}%
where $\mathbf{P}_{(i)}^{(2,2)}$ is the quadrupole-quadrupole polarizability
given through 
\begin{equation}
P_{(i);\alpha ,\beta }^{(2,2)}=\beta _{(i)}(\omega )\bigtriangleup _{\alpha
,\beta }^{(2)}\text{,}  \label{P_1_2}
\end{equation}%
with $\beta _{(i)}(\omega )=\beta _{(i)}^{\shortmid }(\omega )+i\beta
_{(i)}^{\shortparallel }(\omega )$ and 
\begin{eqnarray}
\bigtriangleup _{\alpha ,\beta }^{(2)} &=&\frac{1}{2}\left( \delta _{\alpha
_{1}\beta _{1}}\delta _{\alpha _{2}\beta _{2}}+\delta _{\alpha _{1}\beta
_{2}}\delta _{\alpha _{2}\beta _{1}}\right) b  \notag \\
&&-\frac{1}{3}\delta _{\alpha _{1}\alpha _{2}}\delta _{\beta _{1}\beta _{2}}%
\text{.}  \label{tetra}
\end{eqnarray}%
is the isotropic tetradric. Thus, with Eqs. (\ref{P_1_2}) and (\ref{tetra}),
Eq. (\ref{FD_qua}) is written as 
\begin{equation}
\left\langle M_{(i);\alpha }^{(2)\ast }M_{(i);\beta }^{(2)}\right\rangle =%
\frac{2\varepsilon _{0}}{\pi \omega (c_{2})^{2}}\beta _{(i)}^{\shortparallel
}(\omega )\Theta (\omega ,T_{i})\bigtriangleup _{\alpha ,\beta }^{(2)}\text{.%
}  \label{qua_cor}
\end{equation}%
Up to the quadrupolar order one has to take into account also the cross
correlation dipole-quadrupole 
\begin{equation}
\left\langle \mathbf{M}_{(i)}^{(1)\ast }\mathbf{M}_{(i)}^{(2)}\right\rangle =%
\frac{-i\varepsilon _{0}}{\pi \omega c_{2}}\left( \mathbf{P}_{(i)}^{(1,2)}-%
\mathbf{P}_{(i)}^{(1,2)\ast }\right) \Theta (\omega ,T_{i})\text{,}
\label{dip_qua}
\end{equation}%
where $\mathbf{P}_{(i)}^{(1,2)}$ is the dipole-quadrupole polarizability,
given through 
\begin{equation}
P_{(i);\alpha ,\beta }^{(1,2)}=\gamma _{(i)}(\omega )\square _{\alpha
_{1},\beta _{1},\beta _{2}}^{(1)}\text{,}  \label{odd}
\end{equation}%
with $\gamma _{(i)}(\omega )=\gamma _{(i)}^{\shortmid }(\omega )+i\gamma
_{(i)}^{\shortparallel }(\omega )$ and 
\begin{equation}
\square _{\alpha _{1},\beta _{1},\beta _{2}}^{(1)}=\varepsilon _{\alpha
_{1},\beta _{1},\beta _{2}}\text{,}  \label{axial}
\end{equation}%
an isotropic skew-symmetric tensor. From Eqs.(\ref{dip_qua})-(\ref{axial})
it follows 
\begin{equation}
\left\langle M_{(i);\alpha }^{(1)\ast }M_{(i);\beta }^{(2)}\right\rangle =%
\frac{2\varepsilon _{0}}{\pi \omega c_{2}}\gamma _{(i)}^{\shortparallel
}(\omega )\Theta (\omega ,T_{i})\varepsilon _{\alpha _{1},\beta _{1},\beta
_{2}}.  \label{cross_cor}
\end{equation}

It must be emphasize that the FDT, Eq. (\ref{FD}), applies whenever the
charge distribution of each particle in the presence of mutual interactions
has reached equilibrium with the heat bath. When this is not the case, in
the non-aged regime~\cite{Groot}, one can still use a similar expression of
the FDT in terms of an effective temperature. This can be done through a
generalized Langevin equation~\cite{kubo}, which takes into account the
heat exchange between the NP$i$ and its thermal bath. Relating the momentum
variance of the NP$i$ with its temperature by the equipartition theorem, one can
obtain the effective temperature ($T_{eff (i)}^{(l,k)}$) through the response
of the system due to fluctuations of the multipolar moments~\cite{lapas}. Thus,
\begin{equation}
T_{eff (i)}^{(l,k)}=T_0+\left(T_0-T_b \right) \left[ \left\langle \mathbf{M}%
_{(i)}^{(l)\ast}\odot \mathbf{M}_{(i)}^{(k)}\right\rangle^2 - 1 \right] 
\text{,}  \label{Teff}
\end{equation}
where $T_0$ stands for the initial temperature of the NP$i$ and $T_b$ is the
bath temperature. This expression shows that when the multipole moments
of the particles are uncorrelated, \textit{i.e.} when both particles equilibrate
independently at two different temperatures, the effective temperature coincides
with that of the bath. This is the situation addressed in this paper.

The effective temperature, defined as that for which the system would
equilibrate, is a parameter measuring the distance to the stationary state
in which both particles reach two different temperatures. It can be also
calculated using a relaxation model~\cite{agusti,agusti2}.

\section{The thermal conductance}

In this section, we will calculate the thermal conductance between the two
NPs in the presence of quadrupolar contributions. To this end, we start by
writing Eq. (\ref{energy_2}) as follows 
\begin{eqnarray}
Q_{i\rightarrow j}&=&\frac{-i\omega\varepsilon_{0}}{4}\sum_{l,k=0}^{%
\infty}c_{l}c_{k}\sum_{\left\{ \beta\right\} }\sum_{,\left\{
\alpha\right\}}\left\langle
M_{(i);\beta}^{(l)\ast}M_{(i);\alpha}^{(k)}\right\rangle \times  \notag \\
&&\left(\mathbf{S}_{(j);\beta,\alpha}^{(l,k)}-\mathbf{S}_{(j);\alpha,%
\beta}^{(k,l)\ast}\right)\text{,}  \label{energy_3}
\end{eqnarray}
where from Eqs. (\ref{green}), (\ref{green3}) and (\ref{kernel}) 
\begin{eqnarray}
\mathbf{S}_{(j);\beta,\alpha}^{(l,k)}&=&\frac{1}{(4\pi\varepsilon_{0})^{2}}%
\sum_{n,m=0}^{\infty}d^{-(n+m+l+k+2)}\times  \notag \\
&&A_{(j)}^{(m,n)}(\omega)B_{\beta,\alpha}^{(l,r)}(\mathbf{\hat{d},}\omega)%
\text{,}  \label{kernel_1}
\end{eqnarray}
with 
\begin{eqnarray}
A_{(j)}^{(m,n)}(\omega)B_{\beta,\alpha}^{(l,k)}(\mathbf{\hat{d},}%
\omega)&=&\sum_{\left\{ \gamma\right\}}\sum_{\left\{ \nu\right\} }
Y_{\beta,\gamma}^{(l,n)}(\mathbf{\hat{d}})P_{(j);\gamma,\nu}^{(n,m)}(\omega)
\times  \notag \\
&&Y_{\nu,\alpha}^{(m,k)}(\mathbf{\hat{d}})\text{.}  \label{A_1}
\end{eqnarray}
Hence, by substituting Eq. (\ref{A_1}) into Eq. (\ref{kernel_1}) and the
resulting equation into Eq. (\ref{energy_3}) we obtain 
\begin{eqnarray}
Q_{i\rightarrow j}&=&\frac{1}{(4\pi)^{3}}\sum_{l,k=0}^{\infty}c_{l}c_{k}%
\sum_{n,m=0}^{\infty}d^{-(n+m+l+k+2)}\times  \notag \\
&& C_{(i\rightarrow j)}^{n,m,l,k}\text{,}  \label{energy_4}
\end{eqnarray}
where 
\begin{eqnarray}
C_{(i\rightarrow j)}^{n,m,l,k} &=& \frac{-i\omega\pi}{\varepsilon_{0}}
\left\{ A_{(j)}^{(m,n)} (\omega)\sum_{\left\{ \beta\right\}
}\sum_{\left\{\alpha\right\}} B_{\beta,\alpha}^{(l,k)} \right. \times  \notag
\\
&& \left. \left\langle
M_{(i);\beta}^{(l)\ast}M_{(i);\alpha}^{(k)}\right\rangle -\text{c.c.}
\right\} \text{.}  \label{C}
\end{eqnarray}
Therefore, from Eq. (\ref{energy_4}) we obtain the net heat flux between
both NPs 
\begin{eqnarray}
Q_{12}&=&Q_{1\rightarrow2}-Q_{2\rightarrow1}  \notag \\
&=&\frac{1}{(4\pi)^{3}}\sum_{l,k=0}^{\infty}c_{l}c_{k}\sum_{n,m=0}^{%
\infty}d^{-(n+m+l+k+2)}\times  \notag \\
&&\left( C_{(1\rightarrow 2)}^{n,m,l,k}-C_{(2 \rightarrow
1)}^{n,m,l,k}\right)\text{.}  \label{total_ener}
\end{eqnarray}

In view of Eq. (\ref{axial}) and the symmetric character of the spherical
surface tensors given through Eq. (\ref{surf_tens}) one can prove from \ Eq.
(\ref{A_1}) that 
\begin{equation}
A_{(j)}^{(2,1)}(\omega )B_{\beta ,\alpha }^{(l,k)}=A_{(j)}^{(1,2)}(\omega
)B_{\beta ,\alpha }^{(l,k)}=0\text{.}  \label{null}
\end{equation}%
Moreover, it can be shown that when $m+n=2p+1$ $(p>n)$, $\mathbf{P}_{(j)}^{(n,m)}$
is proportional to an isotropic skew-symmetric tensor
$\mathbf{\square }^{(p)}$ of order $2p+1$ which satisfies~\cite{hess} 
\begin{equation}
\square _{\mu ,\lambda ,\mu ^{\shortmid }}^{(p)}=-\square _{{\mu }%
^{\shortmid },\lambda ,\mu }^{(p)}\text{,}  \label{skew}
\end{equation}%
Therefore, by symmetry reasons only coefficients $C_{(i\rightarrow
j)}^{n,m,l,k}$ for which $n+m=2q$ and $l+k=2s$, with $q$ and $s$ two
positive integers, contribute to the heat flux. Hence, up to quadrupolar
order we can write from Eq. (\ref{energy_4}) 
\begin{eqnarray}
Q_{i\rightarrow j} &=&\frac{1}{(4\pi )^{3}}\{C_{(i\rightarrow
j)}^{1,1,1,1}d^{-6}+(C_{(i\rightarrow j)}^{1,1,2,2}+\ldots  \notag \\
&&+9C_{(i\rightarrow j)}^{2,2,1,1})d^{-8}+C_{(i\rightarrow
j)}^{2,2,2,2}d^{-10}+\ldots \}\text{.}  \label{energy_5}
\end{eqnarray}

Therefore, from Eqs. (\ref{C}), (\ref{energy_5}), and (\ref{C_1})-(\ref{C_21}%
) we arrive at 
\begin{eqnarray}
Q_{i\rightarrow j} &=&\frac{3}{8\pi ^{3}}\left\{ \alpha
_{(i)}^{\shortparallel }\alpha _{(j)}^{\shortparallel }d^{-6}+\right.  \notag
\\
&&15\left( \alpha _{(j)}^{\shortparallel }\beta _{(i)}^{\shortparallel
}+3\alpha _{(i)}^{\shortparallel }\beta _{(j)}^{\shortparallel }\right)
d^{-8}+  \notag \\
&&\left. 140\beta _{(i)}^{\shortparallel }(\omega )\beta
_{(j)}^{\shortparallel }(\omega )d^{-10}+\ldots \right\} \Theta (\omega
,T_{i})\text{.}  \label{energy_6}
\end{eqnarray}%
and consequently 
\begin{eqnarray}
Q_{12}(\omega ) &=&Q_{1\rightarrow 2}-Q_{2\rightarrow 1}  \notag \\
&=&\frac{3}{8\pi ^{3}}\left\{ \alpha _{(1)}^{\shortparallel }\alpha
_{(2)}^{\shortparallel }d^{-6}+\right.  \notag \\
&&\left. 140\beta _{(1)}^{\shortparallel }(\omega )\beta
_{(2)}^{\shortparallel }(\omega )d^{-10}\right\} \Delta \Theta +  \notag \\
&&\left. \frac{45}{8\pi ^{3}}\left\{ \left( \alpha _{(2)}^{\shortparallel
}\beta _{(1)}^{\shortparallel }+3\alpha _{(1)}^{\shortparallel }\beta
_{(2)}^{\shortparallel }\right) \Theta (\omega ,T_{1})-\right. \right. 
\notag \\
&&\left. \left( \alpha _{(1)}^{\shortparallel }\beta _{(2)}^{\shortparallel
}+3\alpha _{(2)}^{\shortparallel }\beta _{(1)}^{\shortparallel }\right)
\Theta (\omega ,T_{2})\right\} d^{-8}  \label{net_ener_1}
\end{eqnarray}%
where $\Delta \Theta \equiv \left\{ \Theta (\omega ,T_{1})-\Theta (\omega
,T_{2})\right\} $.

When \ NPs are at the same temperature $T$, Eq. (\ref{net_ener_1}) reduces
to 
\begin{equation}
Q_{12}(\omega )=\frac{45}{4\pi ^{3}}\left( \alpha _{(1)}^{\shortparallel
}\beta _{(2)}^{\shortparallel }-\alpha _{(2)}^{\shortparallel }\beta
_{(1)}^{\shortparallel }\right) d^{-8}\Theta (\omega ,T),  \label{net_ener_3}
\end{equation}%
whence since the system is in thermal equilibrium 
\begin{equation}
\int_{0}^{\infty }Q_{12}(\omega )d\omega =0.  \label{net_ener_4}
\end{equation}

In the general case, \textit{i.e.} out of equilibrium, we can linearize Eq. (%
\ref{net_ener_1}) with respect to the temperature diference $\Delta T=
T_{1}-T_{2}$ in order to obtain the conductance given through $%
G_{12}(T_{0})=\partial Q_{12}/\partial \left.{\Delta T}%
\right|_{T_{1}=T_{2}=T_{0}}$. We obtain 
\begin{eqnarray}
G_{12}(T_{0})&=&\frac{3}{8\pi^{3}}\int_{0}^{\infty}\Theta^{\prime}(%
\omega,T_{0})\left\{\alpha_{(1)}^{\shortparallel}\alpha_{(2)}^{%
\shortparallel}d^{-6} + \right.  \notag \\
&&60\left(
\alpha_{(1)}^{\shortparallel}\beta_{(2)}^{\shortparallel}+\alpha_{(2)}^{%
\shortparallel}\beta_{(1)}^{\shortparallel}\right) d^{-8}+  \notag \\
&&\left.
5\beta_{(1)}^{\shortparallel}(\omega)\beta_{(2)}^{\shortparallel}(%
\omega)d^{-10} \right\}d\omega\text{,}  \label{conductance}
\end{eqnarray}
where $T_{0}=(T_{1}+T_{2})/2$ is the average temperature, which corresponds
to the final equilibrium temperature that two bodies would reach when
brought into contact and a heat flow established between them~\cite{adkins}.

In the expression we have obtained for the conductance, we can identify the
following contributions:

\begin{description}
\item[(i)] Dipolar 
\begin{equation}
G_{12}^{dip}(T_{0})=\frac{3}{8\pi^{3}} \left(
\int_{0}^{\infty}\Theta^{\prime}(\omega,T_{0})\alpha_{(1)}^{\shortparallel}%
\alpha_{(2)}^{\shortparallel}d\omega \right)d^{-6}\text{.}  \label{dipolar}
\end{equation}
\end{description}

which coincides with the expression obtained in Ref.~\cite{greffet}.

\begin{description}
\item[(ii)] Quadrupolar 
\begin{eqnarray}
G_{12}^{qd}(T_{0})&=&\frac{1}{2\pi^{3}}\int_{0}^{\infty}\Theta^{\prime}(%
\omega,T_{0}) \left\lbrace 45\left(
\alpha_{(1)}^{\shortparallel}\beta_{(2)}^{\shortparallel}+ \right. \right. 
\notag \\
&&\left. \alpha_{(2)}^{\shortparallel}\beta_{(1)}^{\shortparallel} \right)
d^{-8} +  \notag \\
&&\frac{15}{4}\left.
\beta_{(1)}^{\shortparallel}(\omega)\beta_{(2)}^{\shortparallel}(%
\omega)d^{-10} \right\rbrace d\omega\text{.}  \label{quadru}
\end{eqnarray}
\end{description}

\begin{center}
\begin{figure}[tbp]
\includegraphics[scale=0.8]{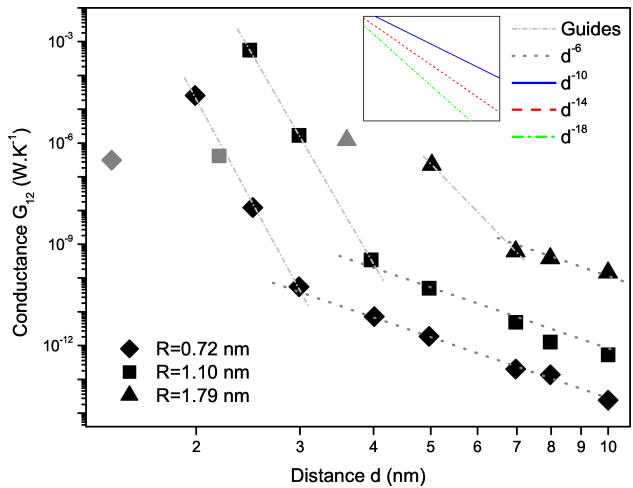}
\caption{Thermal conductance $G_{12}$ vs distance $d$ reproducing the
molecular dynamics data obtained by Domingues \textit{et al.}~\protect\cite%
{greffet}. The grey points represent the conductance when the particles with
effective radius $0.72$, $1.10$, and $1.79$ nanometers are in contact. The
dotted lines show the analytical result obtained by these authors. The
values of the conductance at short distances have been given through the
grey dash-dotted lines. The inset shows different guide behaviors as a
function of the distance having different values of the exponents.}
\label{graph}
\end{figure}
\end{center}

In order to verify our results, in Fig. \ref{graph} we reexhibit a graph
obtained by Domingues \textit{et al.}~\cite{greffet} extending the
logarithmic scale for conductance in the more usual form. This graph
displays the thermal conductance as a function of distance between the NPs,
both with radius $R$, in three significant situations: in mechanical contact
($d=2R$), in the intermediate region shortly before contact ($2R<d<4R$), and
in the most distant region ($d\geqslant 4R$) where the near-field
interaction is still valid. In this situation, the results corresponding to
the grey dotted lines show the behavior $d^{-6}$ which was obtained in Ref.~%
\cite{greffet}. Our results are in broad agreement for this region where the
dipolar domain is present. When the particles are close togheter, their
charge distributions becomes very disorderly and higher orders than dipolar
interactions come into play in the calculation of the thermal conductance.
In this case, as predicted by Domingues et al.~\cite{greffet} the thermal
conductance is about 4 orders of magnitude larger than that of the dipole
model given in Eq. (\ref{dipolar}). In more extreme conditions when the
particles come into contact to each other, the same authors also predicted
that the conductance would be 2 to 3 orders of magnitude lower than the
conductance just before contact. These numerical predictions are covered by
the result we give in Eq.\ (\ref{conductance}) where one can see that the
dominant contribution ($d^{-10}$) is 4 orders of magnitude lower than the
dipolar case ($d^{-6}$) while an intermediate case would give a value $%
d^{-8}.$

It must be stressed that we have obtained the conductance up to quadrupolar
order, nonetheless through our formalism it is possible to obtain the
conductance for any order of multipolar interaction.

\section{Conclusions}

In this paper, we have presented a theory to explain the exchange of energy
between two NPs at different temperature. Our theory provides a general
formalism based on the multipolar expansion of the electrostatic field in
order to study heat transfer between two NPs for arbitrary small distances
provided that the FDT be satisfied. However, out of the FDT regime and when
the system possesses fast and slow degrees of freedom it is possible to
formulate a FDT in terms of a non stationary effective temperature which
depends on the slow degrees of freedom~\cite{agusti,agusti2}.

We have found that our analysis of the heat interchanged between two NPs
separated by a few submicrons agrees with the explains the rapid growth of
the conductance observed in the simulation~\cite{greffet}, even when the NPs
are in contact. Hence, we are able to provide a comprehensive explanation of
the numerical results reported in Ref.~\cite{greffet}.

The formalism presented could also be applied to other situations such as
the radiative heat transfer between a small dielectric particle and a
surface~\cite{greffet4} and the study of the optical forces due the
radiation of a thermal source~\cite{greffet2}, enabling us to go beyond the
dipolar approximation.

\subsection*{Acknowledgments}

One of us (JMR) wants to thank Prof. J.J. Greffet for interesting
discussions. This work was supported by the DGiCYT of Spanish Government
under Grant No. FIS2005-01299, and by Brazillian fellowships CNPq and CAPES.

\renewcommand{\theequation}{A-\arabic{equation}} \setcounter{equation}{0}

\section*{APPENDIX I}

In this Appendix we present some of the properties of the spherical surface
tensors $Y_{\alpha }^{(n)}(\mathbf{\hat{r}})$ related to the $n$-rank
Cartesian tensor 
\begin{equation}
X_{\alpha }^{(n)}(\mathbf{r})=(-1)^{n}\frac{\partial ^{n}}{\partial
a_{\alpha _{1}}\ldots \partial a_{{\alpha _{n}}}}\frac{1}{r}
\label{cart_tens}
\end{equation}%
introduced in Eq. (\ref{multipole}). The tensor $X_{\alpha }^{(n)}(\mathbf{r}%
)$ are given in terms of the unit vector $\mathbf{\hat{r}}$ related to $%
\mathbf{r}$ as~\cite{hess} 
\begin{equation}
X_{\alpha }^{(n)}(\mathbf{r})=r^{-(n+1)}Y_{\alpha }^{(n)}(\mathbf{\hat{r}})%
\text{.}  \label{cart_tens_2}
\end{equation}%
This spherical surface tensors can be expressed as 
\begin{equation}
Y_{\alpha }^{(n)}(\mathbf{\hat{r}})=(2n-1)!!\;\overline{\;\hat{r}_{\alpha
_{1}}\ldots \hat{r}_{\alpha _{n}}}\text{,}  \label{surf_tens}
\end{equation}%
where $\overline{\;\hat{r}_{\alpha _{1}}\ldots \hat{r}_{\alpha _{n}}}$ are
the symmetric irreducible tensor constructed with the components of $\mathbf{%
\hat{r}}$. The first three symmetric irreducible tensors are 
\begin{equation}
\overline{\;\hat{r}_{\alpha }}=r^{-1}r_{\alpha }  \label{example} \\
\end{equation}%
\begin{equation}
\overline{\;\hat{r}_{\alpha _{1}}\hat{r}_{\alpha _{2}}}=r^{-2}\left(
r_{\alpha _{1}}r_{\alpha _{2}}-\frac{1}{3}\delta _{\alpha _{1}\alpha
_{2}}\right)   \label{example_3} \\
\end{equation}%
\begin{eqnarray}
\overline{\;\hat{r}_{\alpha _{1}}\hat{r}_{\alpha _{2}}\hat{r}_{\alpha _{3}}}
&=&r^{-3}\left[ r_{\alpha _{1}}r_{\alpha _{2}}r_{\alpha _{3}}-\frac{1}{5}%
r^{2}\left( \delta _{\alpha _{1}\alpha _{2}}r_{\alpha _{3}}+\right. \right. 
\notag \\
&&\left. \left. \delta _{\alpha _{1}\alpha _{3}}r_{\alpha _{2}}+\delta
_{\alpha _{2}\alpha _{3}}r_{\alpha _{1}}\right) \right] \text{.}
\label{example_2}
\end{eqnarray}%
\textbf{\ }The spherical surface tensors~satisfy the following property%
\textbf{\ }\cite{hess} 
\begin{equation}
\sum_{\left\{ \alpha \right\} }Y_{\alpha }^{(n)}(\mathbf{\hat{r}})Y_{\alpha
}^{(n)}(\mathbf{\hat{r}})=(2n-1)!!n!\text{.}  \label{norm}
\end{equation}

By using Eq. (\ref{A_1}) we can obtain the coefficients defined in Eq. (\ref%
{C}). From Eqs. (\ref{FD_dip})-(\ref{qua_cor}), (\ref{C}), and (\ref{norm})
one has 
\begin{eqnarray}
C_{(i\rightarrow j)}^{1,1,1,1}&=&\frac{-i\omega\pi}{\varepsilon_{0}}%
\{A_{(j)}^{(1,1)}(\omega)\sum_{\left\{ \beta\right\} }
\sum_{\left\{\alpha\right\}} B_{\beta,\alpha}^{(1,1)} \times  \notag \\
&& \left\langle M_{(i);\beta}^{(1)\ast}M_{(i);\alpha}^{(1)}\right\rangle - 
\text{c.c.}\}  \notag \\
&=&4\alpha_{(i)}^{\shortparallel}\alpha_{(j)}^{\shortparallel}\sum_{\left\{%
\beta\right\} }\sum_{\left\{ \nu\right\} }Y_{\beta,\nu}^{(1,1)}
Y_{\beta,\nu}^{(1,1)}\Theta(\omega,T_{i})  \notag \\
&=&24\alpha_{(i)}^{\shortparallel}\alpha_{(j)}^{\shortparallel}\Theta(%
\omega,T_{i})\text{,}  \label{C_1}
\end{eqnarray}
and 
\begin{eqnarray}
C_{(i\rightarrow j)}^{2,2,2,2}&=&\frac{-i\omega\pi}{\varepsilon_{0}}%
\left\{A_{(j)}^{(2,2)}\sum_{\left\{ \beta\right\} }\sum_{\left\{ \alpha
\right\} }B_{\beta,\alpha}^{(2,2)} \times \right.  \notag \\
&&\left. \left\langle
M_{(i);\beta}^{(2)}M_{(i);\alpha}^{(2)\ast}\right\rangle -\text{c.c.}
\right\}  \notag \\
&=&4\beta_{(i)}^{\shortparallel}(\omega)\beta_{(j)}^{\shortparallel}(\omega)%
\sum_{\left\{ \beta\right\} }\sum_{\left\{ \nu\right\}
}Y_{\beta,\nu}^{(2,2)}Y_{\beta,\nu}^{(2,2)}\Theta(\omega,T_{i})  \notag \\
&=&
4(7!!)(4!)\beta_{(i)}^{\shortparallel}(\omega)\beta_{(j)}^{\shortparallel}(%
\omega)\Theta(\omega,T_{i})\text{.}  \label{C_2}
\end{eqnarray}
The remaining coefficients are obtained in similar way 
\begin{eqnarray}
C_{(i\rightarrow j)}^{1,1,2,2}&=&\frac{-i\omega\pi}{\varepsilon_{0}}%
\left\{A_{(j)}^{(1,1)}(\omega)\sum_{\left\{ \beta\right\}
}\sum_{,\left\{\alpha\right\} }B_{\beta,\alpha}^{(2,2)}\times \right.  \notag
\\
&&\left. \left\langle
M_{(i);\beta}^{(2)\ast}M_{(i);\alpha}^{(2)}\right\rangle -\text{c.c.}\right\}
\notag \\
&=&4\alpha_{(j)}^{\shortparallel}(\omega)\beta_{(i)}^{\shortparallel}(%
\omega)\sum_{\left\{ \beta\right\} }\sum_{,\left\{ \alpha\right\}}
Y_{\beta,\nu}^{(2,1)}Y_{\nu,\beta}^{(1,2)}  \notag \\
&=&360\alpha_{(j)}^{\shortparallel}(\omega)\beta_{(i)}^{\shortparallel}(%
\omega)\Theta(\omega,T_{i})  \label{C_12}
\end{eqnarray}
and 
\begin{equation}
C_{(i\rightarrow
j)}^{2,2,1,1}=360\alpha_{(i)}^{\shortparallel}(\omega)\beta_{(j)}^{%
\shortparallel}(\omega)\Theta(\omega,T_{j})\text{.}  \label{C_21}
\end{equation}

\renewcommand{\theequation}{B-\arabic{equation}} \setcounter{equation}{0}

\section*{APPENDIX II}

This Appendix is devoted to the derivation of the expression of the energy
dissipated corresponding to Eq. (\ref{energy}). In the adiabatic case, 
for a perturbation of the form 
\begin{equation}
\hat{H}=-c_{j}\hat{x}_{j}f_{j}(t)\text{,}  \label{H}
\end{equation}%
where $\hat{x}_{j}$ is a generalized displacement and $f_{j}(t)$ is a
generalized force, the change in the energy of the system is equal to the
mean value of the partial derivative of the Hamiltonian with respect to
time. Since only the perturbation $\hat{H}$ in the Hamiltonian depends
explicitely on time and $\hat{x}_{j}$ is a dynamical observable of the
system which is independent of time, we have 
\begin{equation}
dE/dt=-c_{j}x_{j}df_{j}/dt.  \label{energy_change}
\end{equation}%
In the framework of linear response theory one assumes that 
\begin{equation}
c_{j}x_{j}(t)=\int_{0}^{\infty }\alpha _{jk}(\tau )f_{k}(t-\tau )d\tau \text{%
,}  \label{l_r}
\end{equation}%
a relation similar to Eq. (\ref{lin_res}). After introducing the Fourier
transforms and combining Eqs. (\ref{l_r}) and (\ref{energy_change}) we can
write 
\begin{eqnarray}
\frac{dE}{dt} &=&\frac{1}{(2\pi )^{2}}\int d\omega \int d{\omega }^{\prime
}\exp ({-i(\omega +{\omega }^{\prime })t})\times  \notag  \label{heat_1} \\
&&(i{\omega }^{\prime })f_{j}({\omega }^{\prime })\alpha _{jk}(\omega
)f_{k}(\omega )\text{.}
\end{eqnarray}%
Here, if the perturbation $\mathbf{f}$ acts over a finite time, the total
energy dissipated is 
\begin{eqnarray}
\int_{-\infty }^{\infty }dt\frac{dE}{dt} &=&\frac{2}{2\pi }\int d\omega \int
d{\omega }^{\prime }\delta (\omega +{\omega }^{\prime })(i{\omega }^{\prime
})\times  \notag \\
&&f_{j}({\omega }^{\prime })\alpha _{jk}(\omega )f_{k}(\omega )  \notag \\
&=&\frac{1}{2\pi }\int d\omega (-i\omega )f_{j}(-\omega )\alpha _{jk}(\omega
)\times  \notag \\
&&f_{k}(\omega )\text{.}  \label{heat_2}
\end{eqnarray}%
Since the total heat must be a real quantity 
\begin{eqnarray}
\lefteqn{\frac{1}{2\pi }\int d\omega (-i\omega )f_{j}(-\omega )\alpha
_{jk}(\omega )f_{k}(\omega )=}  \notag \\
&&{}-\frac{i}{4\pi }\int d\omega \left( f_{j}^{\ast }\alpha
_{jk}f_{k}-f_{j}\alpha _{jk}^{\ast }f_{k}^{\ast }\right) \omega \text{.}
\label{heat_3}
\end{eqnarray}%
Therefore, the heat at the frequency $\omega $ is given through 
\begin{equation}
Q(\omega )=-\frac{i\omega }{4\pi }\left( f_{j}^{\ast }\alpha
_{jk}f_{k}-f_{k}\alpha _{kj}^{\ast }f_{j}^{\ast }\right) \text{,}
\label{heat_4}
\end{equation}%
which after performing the thermal average leads to the equation equivalent
to Eq. (\ref{energy}) 
\begin{equation}
Q(\omega )=-\frac{i\omega }{4\pi }\left( \langle f_{j}^{\ast }\alpha
_{jk}f_{k}\rangle -\langle f_{k}\alpha _{kj}^{\ast }f_{j}^{\ast }\rangle
\right) \text{.}  \label{heat_5}
\end{equation}%
.

\end{document}